\documentclass[sn-mathphys,Numbered,iicol]{sn-jnl}
\usepackage{graphicx}%
\usepackage{multirow}%
\usepackage{amsmath,amssymb,amsfonts}%
\usepackage{amsthm}%
\usepackage{mathrsfs}%
\usepackage[title]{appendix}%
\usepackage{xcolor}%
\usepackage{textcomp}%
\usepackage{manyfoot}%
\usepackage{booktabs}%
\usepackage{algorithm}%
\usepackage{algorithmicx}%
\usepackage{algpseudocode}%
\usepackage{listings}%
\usepackage{float}%
\usepackage{soul} %
\usepackage{ulem} %
\begin{document}

\title{Efficient Machine Learning Force Field for Large-Scale Molecular Simulations of Organic Systems}

\author[1,2]{\fnm{Junbao} \sur{Hu}}
\author*[3]{\fnm{Liyang} \sur{Zhou}}\email{fhzly@juhua.com}
\author*[1,2]{\fnm{Jian} \sur{Jiang}}\email{jiangj@iccas.ac.cn}

\affil[1]{\orgdiv{Beijing National Laboratory for Molecular Sciences}, \orgname{State Key Laboratory of Polymer Physics and Chemistry}, \orgaddress{\street{Institute of Chemistry, Chinese Academy of Sciences}, \city{Beijing}, \postcode{100190}, \country{P. R. China}}}

\affil[2]{\orgdiv{University of Chinese Academy of Sciences}, \orgaddress{\city{Beijing}, \postcode{100049}, \country{P. R. China}}}

\affil[3]{\orgdiv{Juhua Group Co., Ltd}, \orgaddress{\city{Quzhou}, \postcode{324004}, \country{P. R. China}}}

\abstract{To address the computational challenges of ab initio molecular dynamics and the accuracy limitations of empirical force fields, the introduction of machine learning force fields has proven effective in various systems including metals and inorganic materials. However, in large-scale organic systems, the application of machine learning force fields is often hindered by impediments such as the complexity of long-range intermolecular interactions and molecular conformations, as well as the instability in long-time molecular simulations.Therefore, we propose a universal multiscale higher-order equivariant model combined with active learning techniques, efficiently capturing the complex long-range intermolecular interactions and molecular conformations. Compared to existing equivariant models, our model achieves the highest predictive accuracy, and magnitude-level improvements in computational speed and memory efficiency. In addition,  a bond length stretching method is designed to improve the stability of long-time molecular simulations. Utilizing only 901 samples from a dataset with 120 atoms, our model successfully extends high precision to systems with hundreds of thousands of atoms. These achievements guarantee high predictive accuracy, fast simulation speed, minimal memory consumption, and robust simulation stability, satisfying the requirements for high-precision and long-time molecular simulations in large-scale organic systems.}

\maketitle

\section*{Main}\label{sec:Main}

Molecular dynamics (MD) simulation has gained significant attention in recent years across various disciplines, spanning physics, chemistry, biology, and materials science. This cutting-edge technology, by simulating interactions between molecules or atoms, offers researchers a means to investigate the microstructure of substances and understand their macroscopic properties. This technique has proved invaluable for experimental design, development of new materials, and advances in biomedical research \cite{zhang2018molecular, zhong2023phase, ma2023long, perilla2017physical}. 

Traditional molecular simulations grapple with the dilemma of balancing the high computational cost of ab initio molecular dynamics (AIMD) against the low precision of empirical force fields. A resolution to this challenge is found in the application of machine learning, leveraging its powerful fitting capabilities. The fundamental idea of machine learning force field (MLFF) is to establish a mapping from molecular coordinates to the labels of high-precision quantum chemistry data, including potential energy and forces. This approach eliminates the need to solve the intricate Schrödinger equation, resulting in a significant acceleration and achieving a balance between prediction precision and simulation speed \cite{mouvet2022recent}.

Since the advent of BPNN in 2007\cite{behler2007generalized}, numerous MLFF models have been proposed to improve prediction accuracy, and their performance has been systematically investigated in public datasets (MD17\cite{doi:10.1126/sciadv.1603015}, MD22\cite{chmiela2023accurate}, OC22\cite{tran2023open}). From the perspective of tensor order (denoted by $l$), existing 3D molecular representation learning models can be categorized into two main classes: one is the invariant graph neural networks with only scalar features (i.e., $l=0$) , including SchNet\cite{schutt2017schnet}, DeePMD\cite{wang2018deepmd}, DTNN\cite{schutt2017dtnn}, PhysNet\cite{unke2019physnet}, ComENet\cite{wang2022comenet}, SphereNet\cite{liu2021spherenet}; the other is the equivariant graph neural networks with vector features (i.e., $l=1$) including EGNN\cite{satorras2021egnn}, PaiNN\cite{schutt2021equivariant}, GVP-GNN\cite{jing2020GVP-GNN}, EQGAT\cite{le2022EQGAT}, as well as higher-order features (i.e., $l>1$) such as TFN\cite{thomas2018tfn}, Cormorant\cite{anderson2019cormorant}, SEGNN\cite{brandstetter2021segnn}, Equiformer\cite{liao2022equiformer}, NequlP\cite{batzner20223}, Allegro\cite{musaelian2023allegro}, BotNet\cite{Batatia2022botnet}, MACE\cite{batatia2022mace}. Specifically, invariant methods directly use invariant geometric features such as distance and angles as input, ensuring invariance to the rotation and translation transformations on input molecules. In contrast, the equivariant model can maintain certain symmetries or properties under specific transformations such as rotation and translation.\cite{zhang2023book}. Models with equivariance properties for $l\ge1$ generally outperform invariant ones on various public datasets and tests\cite{satorras2021egnn, batzner20223, kovacs2023mace-eval, fu2022forcesnotenough}. Furthermore, in terms of the expressive power of geometric graph neural networks, higher-order equivariant models demonstrate more expressive capabilities compared to the first-order equivariant models\cite{joshi2023expressive}. Increasing the equivariant order often leads to improved model accuracy\cite{geiger2022e3nn, batatia2022mace, rackers2023recipe}. Therefore, higher-order equivariant force field models such as NequIP, Allegro, and MACE  achieve excellent performance on multiple MD simulation metrics \cite{fu2022forcesnotenough} with nearly a 1000-fold improvement in data efficiency \cite{batzner20223, musaelian2023allegro}.

The high accuracy MLFFs (the errors in atom energy and forces are around 1 meV/atom and 50 meV/\AA, respectively \cite{tokita2023tutorial,klicpera2021gemnet}) have been successfully applied in diverse systems such as metallic and non-metallic inorganic materials\cite{zeng2023deepmdpack}. However, there is limited research to date on the application of MLFF in organic systems. In fact, due to the weak generalization ability of machine learning models\cite{gao2020torchani}, the application of MLFF in large-scale organic systems is usually challenged by several impediments such as the complexity of long-range intermolecular interactions and molecular conformations, as well as the instability in long-time molecular simulations. Therefore, molecular simulations based on MLFF often fail (e.g., loss of accuracy, bond breaking, and atomic overlap) when encountering scenarios not present in the training set \cite{fu2022forcesnotenough,wang2023improving} .

Expanding the receptive field of neural network models by appropriately increasing the number of interaction layers has been proven to enhance the predictive accuracy of MLFF in multi-molecular interaction systems\cite{li2023longMSRA}. 
However, increasing the number of interaction layers results in higher computational costs and can lead to over-smoothing issues that diminish the expressive capability of the machine learning model\cite{di2023over}. In fact, expanding the model's receptive field by increasing the number of interaction layers does not guarantee accurate characterization of long-range intermolecular interactions\cite{kovacs2023mace-eval}. 

The key to accurately capturing long-range intermolecular interactions is increasing the cutoff radius of the neural network model's hyperparameter. However, the number of neighboring atoms increases cubically with increasing cutoff radius, resulting in a significant increase in simulation time and memory consumption. This limitation largely hampers the application of MLFF in long-time MD simulations for large-scale organic systems. Consequently, many models adopt modular methods to consider long-range intermolecular interactions.
For instance, a fragment-based method\cite{li2023longMSRA} was used to demonstrate the long-range intermolecular interactions. However, this fragment-based approch lacks generality and may require specific fragmentation schemes for certain molecules.
The DeePMD-LR method\cite{zhang2022deeplr} directly incorporates existing empirical equations to account for the long-range intermolecular interactions, but the limited applicability and the requirement for expert knowledge in empirical force fields restrict the use of this method.
The fourth generation HDNNP network\cite{ko2021fourth} employs a separate neural network module to predict long-range interactions, which results in an overly complex model with significant computational costs, limiting its application in long-time MD simulations.
The MFN model\cite{batatia2023equivariant} considers the long-range intermolecular interactions by adopting a new architecture that incorporates analytic matrix-valued functions. However, the computational complexity and the memory consumption scale as $O(N^2)$ and $O(N^{4/3})$, respectively, where N is the number of atoms.
Therefore, there is a pressing need for an efficient end-to-end method to capture the long-range intermolecular interactions in the organic systems.

In addition, the extensive conformational variability of organic compounds requires that the training set ensures not only quantity but also diversity\cite{zuo2020performance}. Unfortunately, AIMD proves to be prohibitively expensive to generate a large number of diverse conformations. Consequently, an efficient method for collecting training sets becomes imperative, emphasizing the need for non-redundant datasets to mitigate the labeling costs associated with quantum-mechanical calculations such as density functional theory (DFT).
Moreover, the impracticality of preparing the training set from high-precision quantum DFT calculations for large organic systems further complicates the labeling process. As a result, the labeled datasets are typically derived from smaller systems\cite{huang2021dpgen-eg}. This limitation underscores the necessity of investigating whether MLFF methods can maintain high-precision generalization when applied to larger systems.

Finally, to address the issue of instability in long-time molecular simulations, various methods have been proposed. For example, directly incorporating configurations not present in the training set can enhance the stability of the simulation \cite{wang2023improving}, but it is impractical to exhaustively consider all possible configurations. 
Reducing the time step in MD is another strategy to decrease the incidence of simulation crashes \cite{batzner20223,musaelian2023allegro}. However, this approach comes at the cost of a proportional increase in simulation time.
Although larger models with millions or more parameters may alleviate the issues mentioned above, they do not guarantee simulation stability \cite{fu2022forcesnotenough}.
Therefore, a more efficient and straightforward MLFF model is required to ensure stability during long-time MD simulations.

To address the aforementioned challenges, we propose an end-to-end, highly efficient, and broadly applicable higher-order equivariant model to effectively handle the long-range intermolecular interactions. This method exhibits excellent competitiveness in various aspects, including prediction accuracy, simulation speed, memory consumption, and MD results. Moreover, the incorporation of a bond length stretching technique significantly boosts the simulation stability of MLFF models in organic systems. In addition, an active learning approach based on committee queries is adopted to efficiently collect datasets. Notably, this study successfully extends high precision from a limited dataset comprising 901 samples with 120 atoms to larger systems encompassing ten thousand atoms. This achievement plays a pivotal role in facilitating long-time MD simulations in large-scale organic systems.

\section*{Results}\label{Results}

\subsection*{Efficient and universal multiscale higher-order equivariant modeling architecture}\label{subsec1}

\begin{figure*}[h]
\centering
\includegraphics[width=1.0\linewidth]{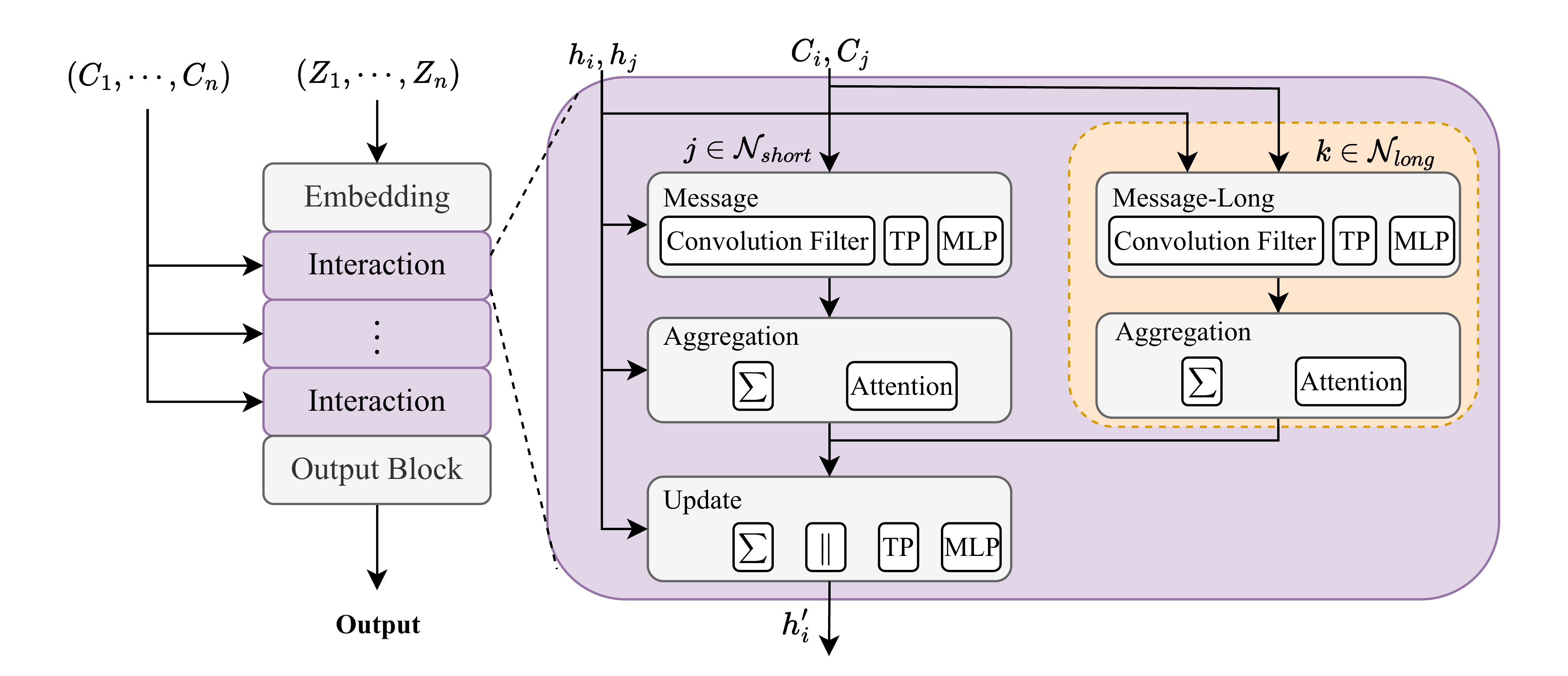}
\caption{\textbar\textbf{The framework of the universal multiscale higher-order equivariant model.}
The model comprises an embedding layer, an interaction layer, and an output layer. The interaction layer may consist of multiple layers that are responsible for considering interactions between the central atom and its neighboring atoms. The yellow module is designed to efficiently capture the long-range intermolecular interactions. 
This module achieves efficient processing of long-range message by reducing the channel numbers of nodes and lowering expansion order of direction information. 
After the aggregation of the long-range message, the feature dimensions of the long-range nodes and the local nodes are aligned, and then they are added together.
Here, \texttt{$\sum$} denotes summation, \texttt{$\parallel$} represents feature concatenation, \texttt{TP} denotes tensor product, \texttt{MLP} stands for multilayer perceptron, and \texttt{Attention} indicates the attention mechanism. 
}
\label{fig:model} 
\end{figure*}

Using the Quantum Mechanics (QM) method to compute large-scale systems remains a challenging task\cite{chung2015oniom,collins2015energy}. However, certain studies focus only on specific regions within the system, such as the active sites of enzymes. 
Therefore, the Quantum Mechanics/Molecular Mechanics (QM/MM) hybrid method, also known as a multiscale method, is widely employed to address this challenge.
The key concept of this method is that for the target regions, a high computational cost QM method is used, and for nontarget regions, a low computational cost Molecular Mechanics (MM) method is employed; thereby achieving efficient computations for large-scale systems\cite{collins2015energy}.

Most of the MLFF models are based on the locality hypothesis of atoms\cite{grisafi2019incorporating, fedik2022extending}. This assumption claims that atoms only interact with their neighbors and is physically justified by the nearsightedness principle of electronic matter. \cite{doi:10.1073/pnas.0505436102} Furthermore, this assumption infers that interactions between atoms that are far apart are negligible. 
In order to take into account long-range intermolecular interactions, the MLFF models must increase receptive field. In this work, inspired by the QM/MM method, we propose a universal multiscale higher-order equivariant model, in which a high-cost module is used to demonstrate the short-range interactions related to the central atom, while a low-cost module is employed to capture the corresponding long-range interactions. 
This method adopts a multiscale strategy that empowers MLFF models to attain optimal precision and efficiency in handling intermolecular long-range interactions.
Specifically, the potential energy of each atom is divided into short-range and long range parts, and the total potential energy of the system is obtained by summing the contributions from all atoms.
The atomic forces are the negative gradients of the total potential energy with respect to the coordinates, ensuring the energy conservation of the model\cite{chmiela2019sgdml}:

$$E_{pot}=\sum_{i\in N_{\text{atoms}}}(E_{i}^{\text{short}}+E_{i}^{\text{long}})$$
$$\overrightarrow{F_i}=-\triangledown_iE_{pot}$$

As mentioned above, in terms of fitting atomic potential energy, higher-order equivariant models significantly outperform the first-order equivariant models and invariant models. Therefore, in this work, we only focus on higher-order equivariant models. Various higher-order equivariant models have been proposed through the construction of different nonlinear functions, convolution filters, message functions, aggregation functions, and update functions. Examples of these models include TFN \cite{thomas2018tfn}, Equiformer \cite{liao2022equiformer}, Cormorant \cite{anderson2019cormorant}, MACE \cite{batatia2022mace}, NequIP \cite{batzner20223nequip}, SEGNNs \cite{brandstetter2021geometric} and so on.
Our universal and efficient multiscale higher-order equivariant model is proposed based on these groundworks (Fig. \ref{fig:model}).
This multiscale model achieves fast processing of long-range messages by reducing the channel dimensions of nodes through equivariant linear transformations and lowering the expansion order of the embedding  direction information.
After the long-range messages are aggregated, another linear transformation is used to align the long-range and short-range features.  
Finally, the summation of these two features gives the entire multiscale equivariant model.  
We have to note that this multiscale model is general. Any higher-order equivariant model can be combined with this model. 
In this work, our multiscale higher-order equivariant model  is constructed based on the MACE model. Hereafter, it will be referred to as ``Our'' or ``Our (MS-MACE)''.

\subsection*{Intermolecular interaction}\label{subsec2}

\begin{figure}
\centering
\includegraphics[width=1.0\columnwidth]{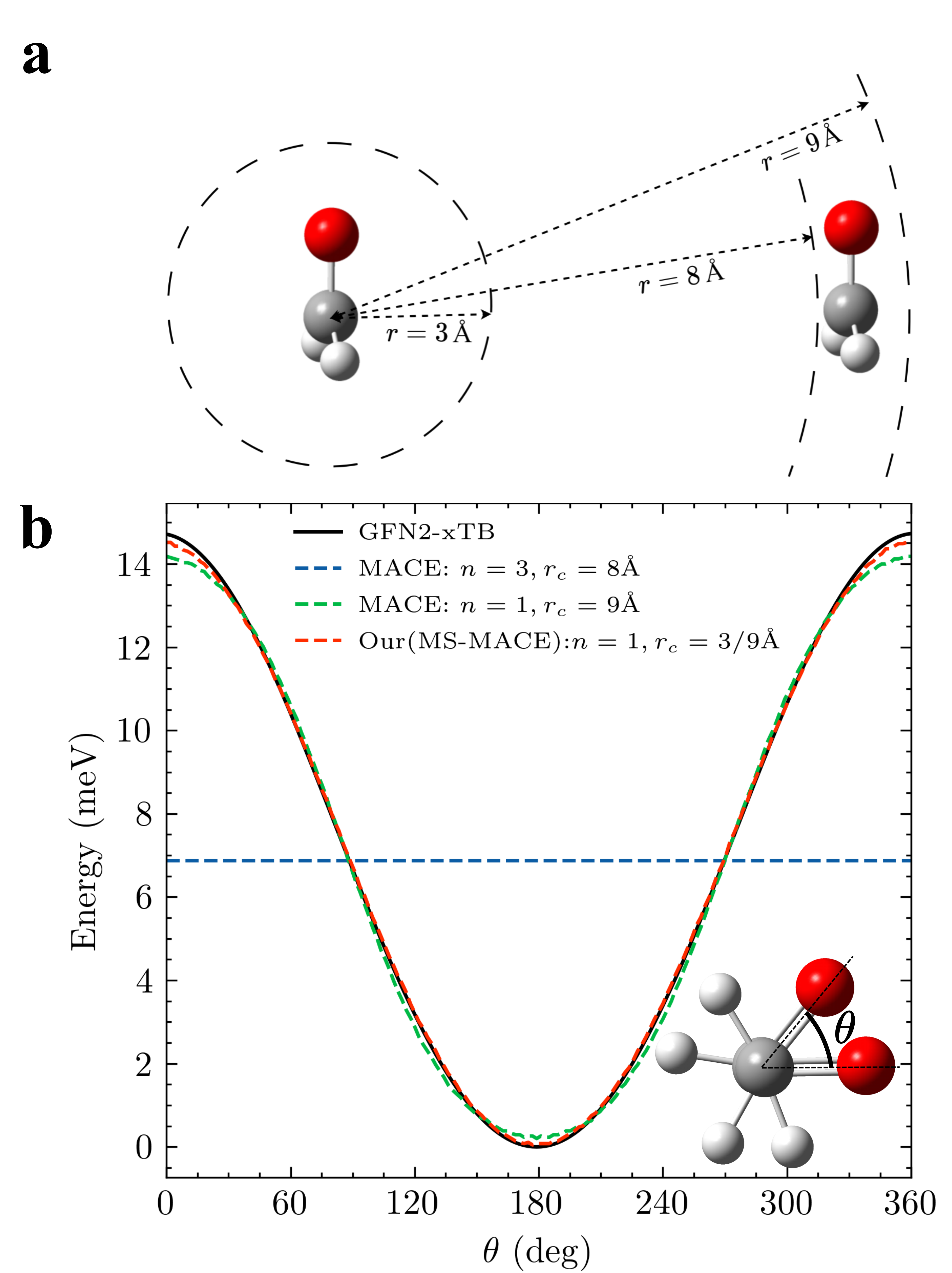}
\caption{\textbar\textbf{The potential energy predictions from three different machine learning models.} 
\textbf{a} The schematic diagram of two formaldehyde molecules. The shortest and farthest distances between the right molecule and the carbon atom of the left molecule are 8 \AA\ and 9 \AA, respectively.
\textbf{b} the potential energy predictions on the test dataset. The blue dash line denotes the model with three interaction layers (n=3) and a cutoff radius of 8 \AA, the green dash line represents the model with one interaction layer (n=1) and a cutoff radius of 9 \AA, the red dash line represents our multiscale model, and the black line represents the GFN2-xTB results.}
\label{fig:Intermolecular} 
\end{figure}

Increasing the number of interaction layers generally enhances the receptive field of the model \cite{lubbers2018hierarchical}. 
However, for long-range intermolecular interactions, there may be scenarios that lack intermediary atoms, which causes an interruption in propagation, consequently leading to the ineffectiveness of this model.
On the contrary, increasing the cutoff radius of the neural network model’s hyperparameter proves to be the key to accurately capturing long-range intermolecular interactions.
For example, we consider the total potential energy of two formaldehyde molecules that are arranged in a face-to-face parallel alignment in space.
Taking the carbon atom of the left molecule as the center atom, the shortest and longest distances to the right molecule are 8\AA\ and 9 \AA, respectively, as shown in \text{Fig. }\ref{fig:Intermolecular}a. 
To prepare the dataset, we rotate one of the molecules around the central axis of two carbon atoms by 1 degree each time, and collect 360 samples. 
In this dataset, we select one sample every 3 degrees to include in the training dataset, while the remaining configurations serve as the test dataset.
The energy and forces for each sample are calculated using the semi-empirical \text{GFN2-xTB} level of theory\cite{bannwarth2019gfn2}.
Three different machine learning models based on the MACE model are considered: (i) three interaction layers (n = 3) with a cutoff radius of 8 \AA; (ii) one interaction layer (n = 1) with a cutoff radius of 9 \AA; (iii) our multiscale model with one interaction layer (n = 1) and cutoff radii of 3 \AA\ for short-range interactions and 9 \AA\ for long-range interactions.
To guarantee sufficient fitting capacity for all models, in the first two models and the short-range part of the third model, the node features are set to \texttt{256x0e}, the direction information is embedded using a third-order expansion, the cutoff polynomial envelope is set to 60, and the batch size is set to 2. In the long-range part of the third model (our MS-MACE model), the node features are set to \texttt{32x0e} with only first order expansion for direction information embeddings. All other hyperparameters remain the same.

Due to the lack of intermediary atoms between these two formaldehyde molecules, the transmission of the message is interrupted. Consequently, merely increasing the number of interaction layers without enlarging the cutoff radius is insufficient to capture long-range intermolecular interactions.  Therefore, according to Fig. \ref{fig:Intermolecular}(b), the first model (n = 3, $r_c$ = 8 \AA) can only predict the average energy (the blue dash line).
On the other hand, although the number of interaction layer is reduced to one (n = 1),  the variations of the potential energy with respect to angles can be effectively captured by increasing the cutoff radius to 9 \AA\ (the green dash line). However, a longer cutoff radius incurs significant computational cost and memory consumption, especially for large-scale systems.
In our multiscale model, inspired by the QM/MM method, we consider the short-range interactions using a very short cutoff radius ($r_c$ = 3 \AA) but a larger channel dimensions (\texttt{256x0e}) and a higher-order expansion for the direction information, while the long-range intermolecular interactions are captured by a large cutoff radius ($r_c$ = 9 \AA) but a smaller channel dimensions (\texttt{32x0e}) and only first-order expansion for the embedded direction information. According to Fig. \ref{fig:Intermolecular}(b), our model gives the most accurate predictions (red dash line). We have to note that our model not only exhibits the highest accuracy, but also achieves magnitude-level improvements in computational speed and memory efficiency.   
That is to say, to consider long-range intermolecular interactions, the MLFF models should increase the cutoff radius rather than the number of interaction layers, the  channel dimensions of node features, and the expansion order for the embedded direction information.  
The accuracy and efficiency of our multiscale model will be discussed in detail below and in Fig. S1 of the Supporting Materials.

\subsection*{Stability of long-time simulation}\label{subsec:3}

\begin{figure*}[htbp]
\centering
\includegraphics[width=1.0\linewidth]{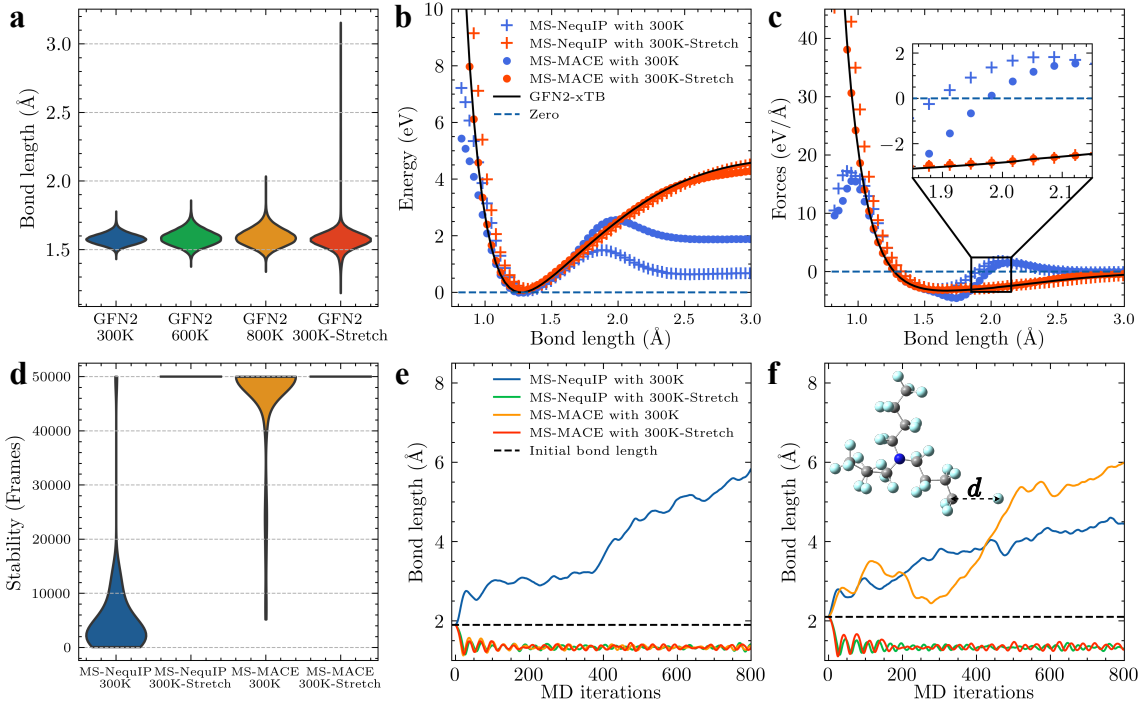}
\caption{\textbar\textbf{Long-time molecular simulation stability.}
\textbf{a}, The violin plots of the C-F bond length distributions for perfluorotri-n-butylamine in the training datasets are collected using GFN2-xTB level of theory at different temperatures. 
In addition, the bond length distribution of the training dataset for 300 K, improved by the bond length stretching method, is also presented (the red shape).
\textbf{b} and \textbf{c} show the average absolute errors of energy and forces for different machine learning models on the test dataset of various C-F bond lengths.
The \texttt{plus} symbol denotes the MS-NequIP model, while the \texttt{dot} symbol denotes the MS-MACE model. The blue color represents the original training dataset derived from the 300 K trajectories (hereafter, this training dataset is referred to as ``300K''), while the red color represents the training dataset improved by the bond length stretching method (hereafter, this training dataset is referred to as ``300K-Stretch''). The black line indicates the GFN2-xTB result and the blue dash line is the zero line.
\textbf{d}, The violin plots for the stability of MD simulations.
In \textbf{e} and \textbf{f}, the curves describe the C-F bond lengths change in MD iterations at 600 K with initial bond lengths of 1.9 \AA\ and 2.1 \AA, respectively. 
The blue and green lines represent the MS-NequIP model trained from the ``300K'' dataset and the ``300K-Stretch'' dataset, respectively. Similarly, the yellow and red lines represent the MS-MACE model trained from the ``300K'' and ``300K-Stretch'' datasets, respectively.}
\label{fig:Stability} 
\end{figure*}

The existing MLFF models are easy to form excessively long or short bonds during long-time MD simulations in organic systems, leading to the collapse of simulations\cite{wang2023improving,stocker2022robust,fu2022forcesnotenough}. 
This is due to the fact that most of the samples in the training dataset are obtained from sampling near the equilibrium state\cite{liu2023discrepancies}], and the proportion of molecular conformations with abnormal bond lengths in the dataset is very low. 
Consequently, the weak generalization ability of machine learning models leads to the collapse of MD simulations.
However, it is challenging to directly collect conformations with extremely long or short bond lengths from MD trajectories, because the MD process would collapse instantaneously in this scenario.
Therefore, we propose a data augmentation method to enhance the stability of the model during long-time MD simulations. We first collect conformations from normal MD trajectories and then use a method of stretching bond lengths to expand the training dataset.

To demonstrate this bond length stretching method, we take an organic molecule as example, i.e.,
perfluorotri-n-butylamine (molecular formula $\text{C}_{12}\text{F}_{27}\text{N}$, CAS No. 86508-42-1), which is a nonconducting, well-thermally and chemically stabilized dielectric liquid and is widely used as a fully immersible electronics coolant.
Molecular simulations of 300 ps are performed at 300 K, 600 K, and 800 K using AIMD simulations based on semi-empirical GFN2-xTB level of theory under the NVT ensemble.
Fig.\ref{fig:Stability}a show the distribution of C-F bond length at different temperatures.
The results indicate that higher temperatures lead to broader distributions.
However, even at 800 K, the distribution of bond lengths remains too narrow to guarantee the stability of long-time MD simulations. Inspired by this fact, the bond length stretching method is developed as follows. First, the training dataset is constructed by randomly selecting 600 samples of conformations from the trajectories at 300 K. 
Then, in every sample within this training dataset, we randomly select a chemical bond and multiply the length of this bond by a factor of $\lambda$, where $\lambda$ is a random variable in the range of [0.85, 2.0]. This range is chosen to avoid overstretching and overcompression of the bonds.
For comparison, the bond length distribution of the dataset with stretched bonds is also added to \text{Fig. }\ref{fig:Stability}a. 
The results indicate that the stretching method can significantly improve the bond length distributions.

To verify the effectiveness of this bond stretching method, we first combine the two representative equivariant models, i.e., the MACE and NequIP models, with our multiscale model, resulting in the MS-MACE and MS-NequIP models, respectively. Then these two multiscale models are trained on datasets that are both processed and unprocessed by the bond length stretching method.
According to Figs. \ref{fig:Stability}b and \ref{fig:Stability}c, our bond length stretching method significantly improves the prediction capabilities of potential energy and forces.  As shown in Fig. \ref{fig:Stability}c, without employing the bond length stretching method, these two MLFF models would even yield unphysical predictions of forces for extremely short or long bond lengths.  
Furthermore, 150 conformations are randomly sampled from the MD trajectories at 800 K to serve as the initial configurations.
Then 50000 steps of Langevin dynamics simulations based on the above MLFF models are performed at 800 K with a time step of 3 fs.
The MD simulation stability of MLFF models can be analyzed by counting the number of frames in which collapse occurs (the criterion for collapse is that there exists a bond length deviating by more than 5 \AA\ from its equilibrium state)\cite{wang2023improving,fu2022forcesnotenough}. 
According to Fig. \ref{fig:Stability}d, without employing the bond length stretching method, the stability of the MS-MACE model is superior to that of the MS-NequIP model, with a success rate of 88.67\%  for the former and only 2.0\% for the latter, further emphasizing the excellent simulation stability of the MACE framework \cite{kovacs2023mace-eval}. 
With employing the bond length stretching method, the simulation stability of both models is greatly enhanced, with a success rate of 100\%.
This suggests that the bond length stretching method exhibits high generality and effectiveness.

\begin{figure*}[htbp]%
\centering
\includegraphics[width=1.0\linewidth]{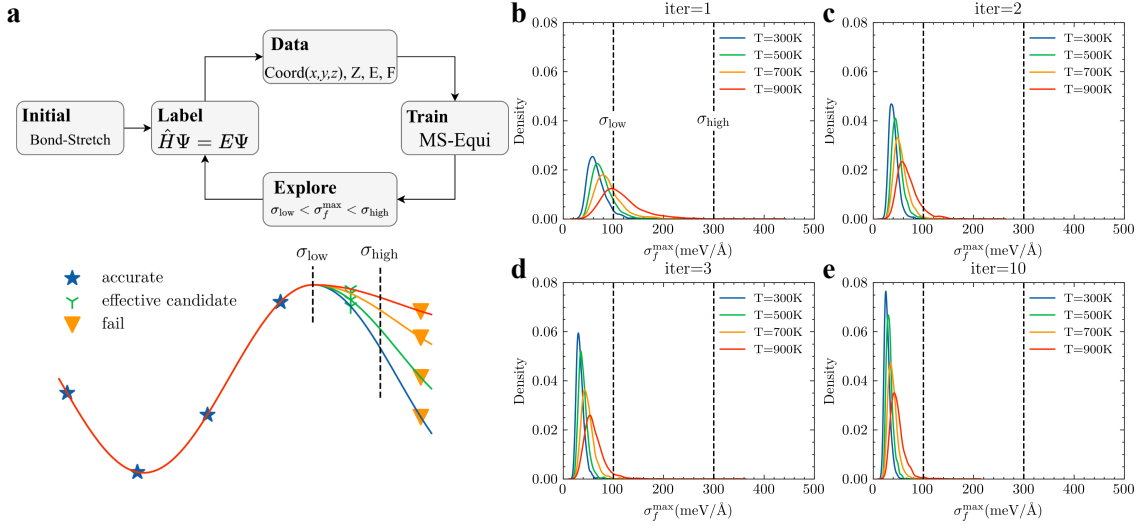}
\caption{\textbar\textbf{Active learning workflow and results.}
\textbf{a} The schematic diagram of active learning techniques within the framework of the multiscale higher-order equivariant model combined with the bond length stretching method.
\textbf{b, c, d} and \textbf{e} respectively show the distribution of the maximum force variance ($\sigma_f^{\text{max}}$) in the first, second, third, and tenth rounds of the data collecting process (or exploring process).
Different colors represent exploration at different temperatures, where the blue, green, yellow, and red lines correspond to 300 K, 500 K, 700 K, and 900 K, respectively.}
\label{fig:active learning } 
\end{figure*}

In addition, according to the inset of Fig. \ref{fig:Stability}c, without employing the bond length stretching method, MS-NequIP and MS-MACE models give unphysical preditions (i.e., repulsive forces) when the bond lengths are larger than 1.9 \AA\ and 2.0 \AA, respectively. This seems to suggest that the threshold bond length for unphysical predictions determines the stability of the long-time MD simulations. Therefore, we infer that the simulation stability of the MLFF model is highly correlated with the predictive capability of extreme bond lengths. 
To further examine the simulation stability of each model, we investigate the change of C-F bond length over the simulation time in Langevin dynamics simulations with a time step of 1 fs at 600 K, where the initial bond lengths are 1.9 \AA\ (Fig. \ref{fig:Stability}e) and 2.1 \AA\ (Fig. \ref{fig:Stability}f), respectively.
According to Fig. \ref{fig:Stability}c, in the case where the bond length is 1.9 \AA\ (larger than the equilibrium bond length), the MS-MACE predicts a negative force (i.e., elastic rebound force). Therefore, the elongated chemical bond can be restored to its equilibrium state (see the yellow line in Fig. \ref{fig:Stability}e). However, in this case, the MS-NequIP model gives an unphysical prediction (a repulsive force) (Fig. \ref{fig:Stability}c), resulting an abnormal elongation of the C-F bond during the MD simulations (see the blue line in Fig. \ref{fig:Stability}e), and ultimately leading to the collapse of simulations. 
At a bond length of 2.1 \AA, without employing  the bond length stretching method, both models give unphysical predictions (i.e., repulsive forces), resulting in incorrect elongations of the C-F bonds (see the yellow and blue lines in Fig. \ref{fig:Stability}f). On the other hand, by the application of our bond length stretching method, both models can make correct predictions for the cases where the initial bond lengths are 1.9 \AA\ and 2.1 \AA, respectively (see the green and red curves in \text{Figs. }\ref{fig:Stability}e and \ref{fig:Stability}f).
In summary, the bond length stretching method proves to be an effective and universally applicable method to enhance the stability of long-time MD simulations.

\subsection*{Efficient collection of training datasets}\label{subsec4}

Due to the limited generalization capability of neural networks, the performance quality of MLFF model is heavily dependent on the quality of the training dataset.
Moreover, labeling the dataset requires the use of quantum-mechanical calculations, which involve a high computational cost.
Therefore, a key task  prior to training MLFF model is to construct a dataset that is as comprehensive and non-redundant as possible \cite{tokita2023tutorial, schutt2020machine}. Active learning techniques have emerged as a prevalent strategy for collecting datasets, with committee querying methods being particularly prominent in the field of MLFF \cite{zhang2020dpgen, huang2021dpgen-eg, fedik2022extending}.

\begin{table*}[t]
\caption{The percentage of accurate ($\sigma_f^{\text{max}} < \sigma_{\text{low}} $), effective candidate ($\sigma_{\text{low}} \le \sigma_f^{\text{max}} \le \sigma_{\text{high}} $), and failed ($\sigma_f^{\text{max}} > \sigma_{\text{high}} $) samples  in the $i^{th}$ iteraction.}\label{al_add}%
\begin{tabular}{lcccccccccc}
\toprule
\textbf{i}   & \textbf{1} & \textbf{2} & \textbf{3} & \textbf{4} & \textbf{5} & \textbf{6} & \textbf{7} & \textbf{8} & \textbf{9} & \textbf{10}\\
\midrule
\textbf{Accurate}     & 73.20 &	97.42 &	98.74 &	98.68 &	98.15 	&99.49 	&99.46 &	99.51 &	99.67 	&99.68  \\
\textbf{Candidate}   & 26.72 	&2.58 &	1.25 &	1.32 	&1.85 &	0.51 	&0.54 &	0.49 &	0.33 	&0.32   \\
\textbf{Failed}   & 0.08 &	0.00 &	0.01 &	0.00 	&0.00 &	0.00 	&0.00 &	0.00 &	0.00 	&0.00  \\
\botrule
\end{tabular}
\end{table*}

The fundamental idea of an active learning technique is to use the disagreement of the committee to quantify the generalization error \cite{schran2020committee}. 
Specifically, if a training dataset is of sufficiently high quality, a MLFF model, operating under various random seed conditions, will yield accurate predictions with low variance for normal samples that are not present in the training set. Otherwise, predictions with high variance indicate that the training dataset is incomplete and requires the addition of more effective samples for further improvement.  
Typically, the maximum variance in force predictions made by the MLFF model under different random seeds is used as a criterion to determine whether a sample should be added to the training set. 
This maximum variance is calculated based on the following formula:
\begin{equation}
\sigma_f^{\text{max}} = \mathop{\text{max}}\limits_{i} \left\{ \sqrt{\langle \Vert  F_i (R_t) - \langle F_i(R_t) \rangle\Vert^2\rangle} \right\}
\label{eq:error_f}
\end{equation}
where $i$ is the index of the atom in the candidate sample. If the maximum variance satisfies the inequality $\sigma_{\text{low}}\le \sigma_f^{\text{max}} \le \sigma_{\text{high}}$, then the sample will be added to the training dataset, where the lower and upper limits are set to $\sigma_{\text{low}}$ = 100 meV/\AA\ and $\sigma_{\text{high}}$ = 300 meV/\AA, respectivley.\cite{huang2021dpgen-eg}
If the maximum variance exceeds the upper limit, it suggests that the corresponding sample may be abnormal. Using it as labeled data could potentially be detrimental to the training effectiveness of the MLFF model. An extremely low maximum variance ($<\sigma_{\text{low}}$) indicates that the existing training dataset is sufficiently comprehensive for the MLFF model to make correct predictions about the corresponding sample. That is to say that this sample does not need to be added to the training dataset.

In this section, similar to the previous section, we will take perfluorotri-n-butylamine molecules as an example to demonstrate the powerful capabilities of active learning technique for data collection within the framework of our multiscale higher-order equivariant model. To construct a dataset  that is as comprehensive and non-redundant as possible, the workflow of the active learning method involves a series of continuous iterations. First, an annealing simulation is conducted using the general AMBER force field (GAFF)\cite{wang2004development} on a system consisting of 200 perfluorotri-n-butylamine molecules in the NPT ensemble. During the annealing simulation, the system cools down from 800 K to 280 K in 60 ns, and 300 frames of trajectory are randomly extracted with a minimum sampling interval of 10 ps. In each frame, the three molecules closest to the center of the simulation box are selected as a sample (120 atoms) and added to the initial dataset. To enhance the simulation stability during data collection, the initial dataset is improved using the bond length stretching method proposed in the previous section. Then the improved dataset is labeled using the semi-empirical GFN2-xTB level of theory.  
Second, four MLFF models based on our multiscale higher-order equivariant model (i.e., MS-MACE) with different random seeds and a batch size of 64 are trained on the initial dataset (300 samples). Then high precision MD simulations are conducted in the NVT ensemble at 300 K, 500 K, 700 K, and 900 K, respectively, using the four trained MLFF models with a step length of 1 fs and a total simulation time of 45 ps. 
In the meantime, a certain number (10, 20, 40, and 50 for 300 K, 500 K, 700 K, and 900 K, respectively.) of new samples are randomly selected as the candidate samples from the MD trajectories. Finally, the candidate samples are assessed using Eq. (\ref{eq:error_f}), and the effective candidates are labeled and added to the training dataset. The above process is repeated until the proportion of effective candidates generated from the pool of candidate samples converges to a value less than 0.5\%.

\begin{figure*}[htbp]
\centering
\includegraphics[width=0.8\linewidth]{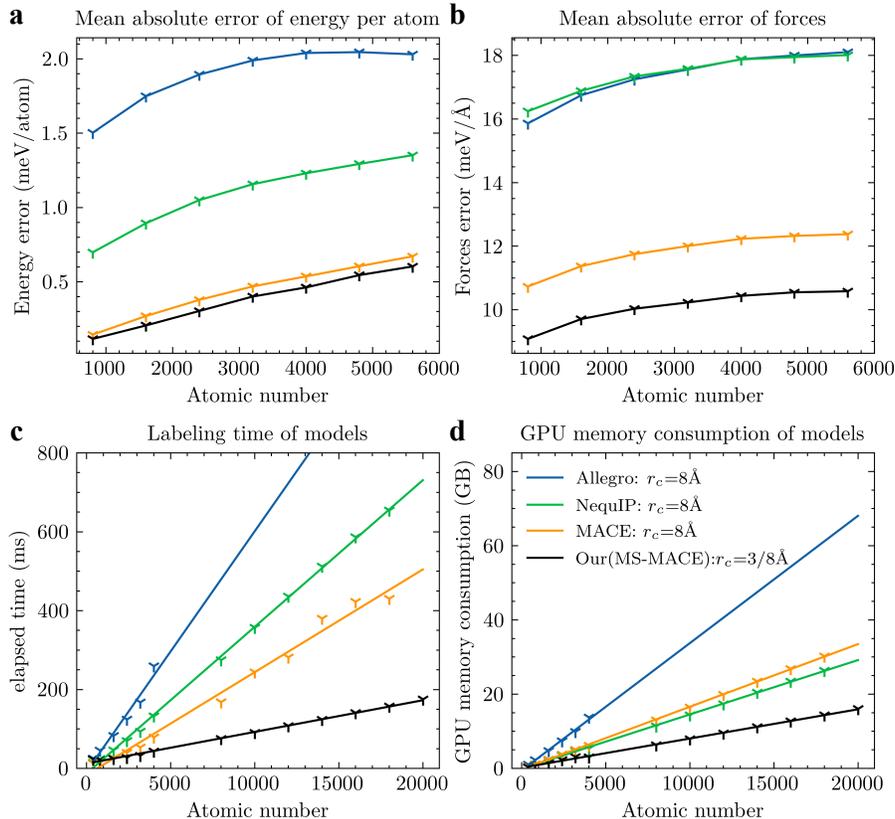}
\caption{\textbar
\textbf{a} and \textbf{b} are the changes of the average absolute errors of the energy and forces per atom with the increase of the atom numbers, respectively. \textbf{c} and \textbf{d} are the changes of time cost and memory consumption of simulations with the increase of atom numbers.
The straight lines are obtained by linear fittings to the data points using the least squares method. The blue, green, yellow, and black lines are from the Allegro, NequIP, MACE, and our MS-MACE models, respectively.}\label{fig:Prediction accuracy} 
\end{figure*}

\text{Fig. 4b-e} depict the distribution of $\sigma_f^{\text{max}}$ at different temperatures in the 1st, 2nd, 3rd, and 10th iterations. \text{Table 1}  presents the percentages of accurate ($\sigma_f^{\text{max}} < \sigma_{\text{low}} $), effective candidate ($\sigma_{\text{low}}\le\sigma_f^{\text{max}} \le \sigma_{\text{high}} $), and failed ($\sigma_f^{\text{max}} > \sigma_{\text{high}} $)  samples in each iteration. In the 1st iteration, 26.80\% of samples exceed the lower limit of the force variance (i.e., $\sigma_f^{\text{max}} \ge \sigma_{\text{low}} $), and the density distribution curves of $\sigma_f^{\text{max}}$ at different temperatures are relatively broad (Fig. \ref{fig:active learning }b). In the first iteration, 26.72\% of the candidates become effective candidates and are added to the initial training dataset.
After just one iteration, the percentage of accurate samples increased to 97.42\%, and the density distribution of $\sigma_f^{\text{max}}$ is primarily concentrated in the range between 0 and $\sigma_{\text{low}}$ (Fig. \ref{fig:active learning }c). The proportion of effective candidate samples drops to 2.58\%.
After 10 steps of iteration, the percentage of accurate samples approaches 99.7\%, and only 0.32\% of samples need to be added to the dataset.
Referring to pertinent literature \cite{zhang2020dpgen,huang2021dpgen-eg}, we conclude that the iterations of data collection using an active learning technique  have converged. In the end, we collect a training  dataset consisting of 901 samples.

\subsection*{Prediction accuracy and molecular simulation efficiency}\label{subsec5}

\begin{figure*}[htbp]%
\centering
\includegraphics[width=1.0\textwidth]{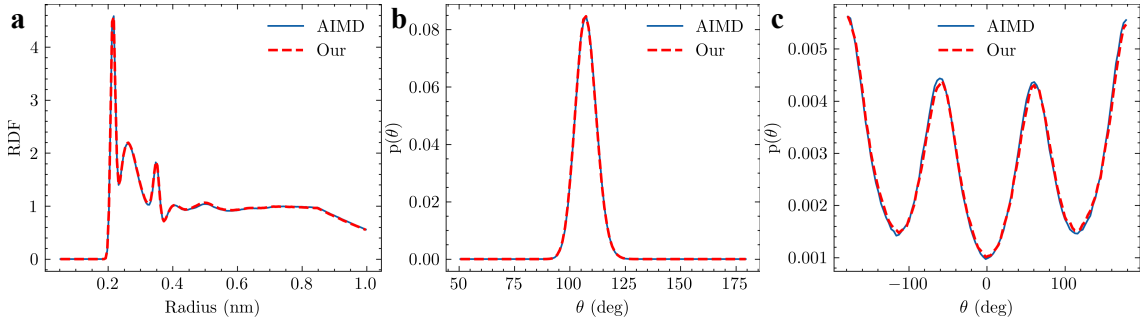}
\caption{\textbar\textbf{Comparisons with AIMD simulations.} 
\textbf{a} is the RDF of F-F atoms.
\textbf{b} is the angle distributions of F-C-F bonds, 
and \textbf{c} is the dihedral distributions of F-C-C-F. 
The blue lines are the results from AIMD simulations and the red dotted lines are the results from our multiscale higher-order equivariant model.}
\label{fig:rdf} 
\end{figure*}

In this section, we compare our multiscale model (referred as ``MS-MACE'' model) with three other representative equivariant models, including MACE, NequIP, and Allegro, in terms of prediction accuracy, simulation speed, and GPU memory consumption. 
It is important to note that achieving complete consistency in hyperparameters for all models is impractical due to differences in model architecture. For example, MACE and NequIP models utilize atoms for message passing, while the Allegro model employs edges for message passing\cite{zhang2023book}. However, to ensure the validity of the comparisons, the shared parameters of the models are set identically. The detailed information on the parameter settings can be found in the Supporting Materials and the section of Method. In addition, it merits emphasis that comparisons with invariant models have not been conducted, because equivariant models typically exhibit superior performance over invariant models in the construction of MLFF.

In the previous section, we have obtained a dataset comprising a total of 901 samples, each containing three perfluorotri-n-butylamine molecules. This dataset is randomly divided into training and validation sets, comprising 801 and 100 samples, respectively.  
Subsequently, we will construct seven test sets, each comprising 100 samples. Within a single test set, each sample contains an equal number of molecules with different conformations. Across the first to the seventh test sets, each sample will contain 20, 40, 60, 80, 100, 120, and 140 molecules, respectively. In order to obtain these seven sets, we conduct all-atom MD simulations on a system with 1000 molecules in the NPT ensemble at 600 K using GAFF force field. All of the samples in the seven test sets are randomly extracted from the MD trajectories, and then labeled using the semi-empirical GFN2-xTB level of theory. In addition, aiming to reduce the impact of randomness, the average prediction errors for each model are obtained from the model with five different random number seeds.

Fig. \ref{fig:Prediction accuracy} shows the results of the comparisons among different equivariant models. All models exhibit high precisions in predicting energy and forces on the test sets. Moreover, the time cost and GPU memory consumption of simulations change linearly with the increase of atom number in the system.
This could be attributed to the fact that all models are based on a local assumption \cite{huguenin2023physics,grisafi2019incorporating,fedik2022extending}.

Compared to other models, our MS-MACE model achieves the best performance across various metrics, including accuracy, simulation speed, and GPU memory consumption. 
Specifically, according to Figs. \ref{fig:Prediction accuracy}a and \ref{fig:Prediction accuracy}b, our MS-MACE model gives the most accurate predictions for both energy and force. In particular, the average error in force approaches an asymptotic value of 10.5 meV/\AA, indicating that even in a large-scale system comprising several hundred thousand atoms, the force error remains around 10.5 meV/\AA. That is to say, although each sample in the training dataset only contains 120 atoms, our multiscale MLFF model can successfully extend high precision to large-scale systems with hundred thousand atoms. In addition, according to Figs. \ref{fig:Prediction accuracy}c and \ref{fig:Prediction accuracy}d, our MS-MACE model achieves magnitude-level improvements in computational speed and memory efficiency. Furthermore, due to the linear characteristics of the time cost and memory consumption changes, it is easy to speculate that when the number of atoms in the system exceeds 100000, the computational speed and memory efficiency will see orders of magnitude improvement.  
In summary, our MS-MACE model demonstrates exceptional competitiveness in energy and force predicting, simulation speed, and GPU memory consumption.

\subsection*{Compared with AIMD simulations}\label{subsec6}

In this section, by comparing with the results from AIMD simulations, we validate the capability of our multiscale higher-order equivariant model. We conduct AIMD simulations in the NVT ensemble at 600 K for 500 ps. The system comprises five perfluorotri-n-butylamine molecules, where periodic boundary conditions are used in all three directions. 
Under the same conditions, MD simulations based on MS-MACE model are performed.  To reduce the influence of randomness, all simulations are repeated three times independently.
Fig. \ref{fig:rdf} shows the comparisons of radial distribution functions (RDFs) for F-F atoms, the angle distributions for F-C-F bonds, and the dihedral distributions for F-C-C-F bonds.
The results indicate that the MS-MACE model can perfectly reproduce the results from AIMD simulations, highlighting the accuracy of our multiscale higher-order equivariant model.

\section*{Discussion}

Although the polarity of the perfluorotri-n-butylamine molecule is extremely weak, a cutoff radius of 8 \AA\ is still required to accurately capture the intermolecular long-range interactions. However, a large cutoff radius is a disaster for the MLFF model. According to Figs. \ref{fig:Prediction accuracy}c and \ref{fig:Prediction accuracy}d, the cost of simulation time and memory consumption is enormous for a general MLFF model, which is unaffordable for simulating large-scale systems. Moreover, the cutoff radius in systems with dipole-dipole or Coulomb interactions will be even larger. Therefore, a multiscale higher-order equivariant model with siginificant advantages in prediction accuracy, simulation speed, and memory consumption is necessary for large-scale organic systems.

In addition, long-time simulation stability is one of the most important and challenging issues in the development of the MLFF model. To address this issue, we proposed a bond length stretching method, in which we randomly select chemical bonds from the training dataset and randomly stretch or compress them within a reasonable range. This method achieves long-time simulation stability comparable to AIMD simulations with almost no additional computational cost. Moreover, this bond length stretching method can effectively improve the long-time simulation stability of all MLFF models.

Finally, a training set that is as comprehensive and non-redundant as possible is also extremely important for the MLFF model, as it directly determines the prediction accuracy and training efficiency of a MLFF model. The active learning technique provides a method for constructing a comprehensive training set; however, active learning is merely a concept. Its effectiveness depends on the corresponding MLFF model. In this work, the active learning concept is combined with the bond length stretching method within the framework of our multiscale higher-order equivariant model. The results show that this method can efficiently and stably construct a comprehensive and low-redundancy training set. For example, we built a training set containing only 901 samples, each with merely 120 atoms. However, based on this dataset, our multiscale model can achieve high-precision, high-efficiency, and low-memory consumption in long-time stable simulations of systems with hundreds of thousands of atoms.

In summary, in this work we proposed a universal multiscale higher-order equivariant framework for constructing a MLFF model. Moreover, within this framework, a bond length stretching method and an active learning workflow have been designed to realize the long-time stability of MD simulations and efficient collection of a comprehensive, low-redundancy training set. By employing our multiscale model, one can achieve high-precision, high-speed, and low GPU memory consumption in long-time stable simulations of large-scale organic systems. In our subsequent work, we will explore the application of this multiscale model in systems with polar or charged molecules.

\section*{Method}\label{Method}

\subsection*{Equivariant model}\label{subsecM1}

Incorporating the data symmetry in machine learning models can improve the efficiency of data collection and the generalization capability of the models \cite{cohen2016group}. For atomic systems, if the coordinates of the system rotate, quantities like forces and dipole moments should rotate accordingly. 
More strictly speaking, if the function $ f : X \rightarrow Y $ is equivariant under the action of a group of transformation $G$, the the following equation holds:
\begin{equation}
f (D_X [g] x) = D_Y [g] f (x).
\label{eq:equivariant}
\end{equation}
In Eq. (\ref{eq:equivariant}), $X$ and $Y$ are two vector spaces, $x$, $y$, and $g$ are elements in $X$, $Y$, and $G$, respectively, and $D_X [g]$ and  $D_Y [g]$ are the transformation matrices parametrized  by g in $X$ and $Y$. 
A natural and effective method to ensure the equivariance of the transformation is to impose constraints on the transformation matrices to consider the symmetric properties of the data.\cite{Batatia2022botnet,darby2023tensor}

Han\cite{han2022} categorizes the equivariant models into three types: vector-based, Lie group-based, and irreducible representation-based. Among these three types of models, the irreducible representation-based approach , which takes advantage of the transformation properties of spherical harmonics $Y_m^l$, exhibits higher-order equivariant capabilities and excels in various force field tasks \cite{fu2022forcesnotenough}. Currently, this method can be unified within the \texttt{e3nn} framework \cite{geiger2022e3nn}.

\subsection*{Multi-scale higher-order equivariant model}\label{subsecM2}

Taking the MACE model as an example, we will illustrate the construction of the multi-scale higher-order equivariant model as shown in Fig. \ref{fig:model}.
First, the atomic numbers are initially mapped to a one-dimensional vector $\delta_{z_{i}}$ through one-hot encoding. Subsequently, a linear transformation initializes the node as $h_{i,c00}^{(0)}$, where the index ``00'' indicates that the current node contains only scalar information.
In the subsequent network framework, node features are denoted by $\bar{h}^{(t,\text{short/long})}_{i,cl_2m_2}$, where $i$ represents the atomic number, $l_2m_2$ signifies the specific spherical harmonic features, $c$ denotes the channel count, $t$ indicates the interaction layer, and ``$\text{short/long}$'' is a shorthand to avoid repetition. 
This is because the operations for short-range and long-range interactions in the interaction layer are similar and implemented through Eqs. (\ref{eq:linear_first_short}) to (\ref{eq:atomic-basis-t}). The key difference lies in the fact that the number of neighboring atoms to handle long-range interactions is larger. 
Therefore, the number of channels $c$ and the order $l$ for long-range interactions should be smaller than those for short-range interactions to reduce computational complexity. Finally, short-range and long-range information is summed up using Eq. (\ref{eq:add}).

Eq. (\ref{eq:linear_first_short}) represents a linear transformation that satisfies equivariance requirements, which only occurs between the same orders. Therefore, it is necessary to specify the current $L$-equivariant information. Eq. (\ref{eq:rad_feats}) denotes the radial embedding function, using Bessel basis functions and polynomial smooth truncation functions.\cite{klicpera2003dimenet} Here, $n$ represents the embedding dimension, and the truncation function for long-range interactions should be greater than that for short-range interactions. Eq. (\ref{eq:radial_MLP}) expands the radial information to a specified dimension using a learnable multilayer perceptron ($\text{MLP}$), which is related to the tensor product in Eq. (\ref{eq:phi-basis-t}).

\begingroup\makeatletter\def\f@size{7.5}\check@mathfonts
\def\maketag@@@#1{\hbox{\m@th\large\normalfont#1}}%
\begin{align}
\label{eq:element_embedding}
h_{i,c00}^{(0)} &=  {\sum_z} W_{cz} \delta_{z_{i}}   \\  
\label{eq:linear_first_short}
\bar{h}^{(t,\text{short/long})}_{i,cl_2m_2} &= \sum_{\tilde{c}} W_{c\tilde{c}l_2}^{(t)} h^{(t)}_{i,\tilde{c}l_2m_2}   \\
\label{eq:rad_feats} 
j^{\text{short/long}}_n (r_{ij}) &=  \sqrt{\frac{2}{r^{\text{short/long}}_{\text{cut}}}} \frac{\sin{\left(n\pi\frac{r_{ij}}{r^{\text{short/long}}_{\text{cut}}} \right)}}{r_{ij}} \nonumber \\
 & \qquad\qquad \times f^{\text{short/long}}_{\text{cut}}(r_{ij})    \\
\label{eq:radial_MLP}
 R_{c \eta_{1} l_{1}l_{2} l_{3}}^{(t,\text{short/long})}(r_{ij}) &=   {\rm MLP}\left( \left\{ {j^\text{short/long}_n} (r_{ij})\right\}\right)  \\
\label{eq:phi-basis-t}
 \phi_{ij,c \eta_{1} l_{3}m_{3}}^{(t,\text{short/long})} &=  \sum_{l_1l_2m_1m_2}C_{\eta_1,l_1m_1l_2m_2}^{l_3m_3} \nonumber \\
 & \times R_{c \eta_{1} l_{1}l_{2} l_{3}}^{(t,\text{short/long})}(r_{ij}) Y^{m_{1}}_{l_{1}} (\boldsymbol{\hat{r}}_{ij}) \bar{h}^{(t,\text{short/long})}_{j,cl_2m_2} \\
\label{eq:atomic-basis-t}
A_{i,cl_{3}m_{3}}^{(t,\text{short/long})} &= \sum_{\tilde{k}, \eta_{1}} W_{c \tilde{c} \eta_{1}l_{3}}^{(t)}
    \sum_{j \in \mathcal{N}(i)}  \phi_{ij,\tilde{k} \eta_{1} l_{3}m_{3}}^{(t,\text{short/long})}  \\
\label{eq:add} 
A_{i,cl_{3}m_{3}}^{(t)} &= ( A_{i,cl_{3}m_{3}}^{(t,\text{short})} + A_{i,cl_{3}m_{3}}^{(t,\text{long})}) / 2 \\
\label{eq:symmbasis_L1} 
  {B}^{(t),\nu}_{i,\eta_{\nu} c LM}
  &= \sum_{{l}{m}} \mathcal{C}^{LM}_{\eta_{\nu}   l   m}  \prod_{\xi = 1}^{\nu} A_{i,c l_\xi  m_\xi}^{(t)}  \\
m_{i,c LM}^{(t)} &=  \sum_{\nu}\sum_{\eta_{\nu}} W_{z_{i} \eta_{\nu} c L}^{(t),\nu} {  B}^{(t),\nu}_{i,\eta_{\nu} c LM}  \label{eq:message}\\
 h^{(t+1)}_{i,c LM}
    &= \sum_{\tilde{c}} W_{c L,\tilde{c}}^{(t)} m_{i,\tilde{c}LM}^{(t)}
  + \sum_{\tilde{c}} W_{z_{i}cL,\tilde{c}}^{(t)}
  h^{(t)}_{i,\tilde{c}LM}   \label{eq:update}\\
E_i &= \sum_{t} \sum_{c}W^{(t)}_{c}h^{(t)}_{i,cLM} \label{eq:site-energy}    \\
 {F} &= -\nabla \sum_{i} E_{i}\label{eq:F}
\end{align}
\endgroup

\begin{figure*}
\centering
\includegraphics[width=0.9\linewidth]{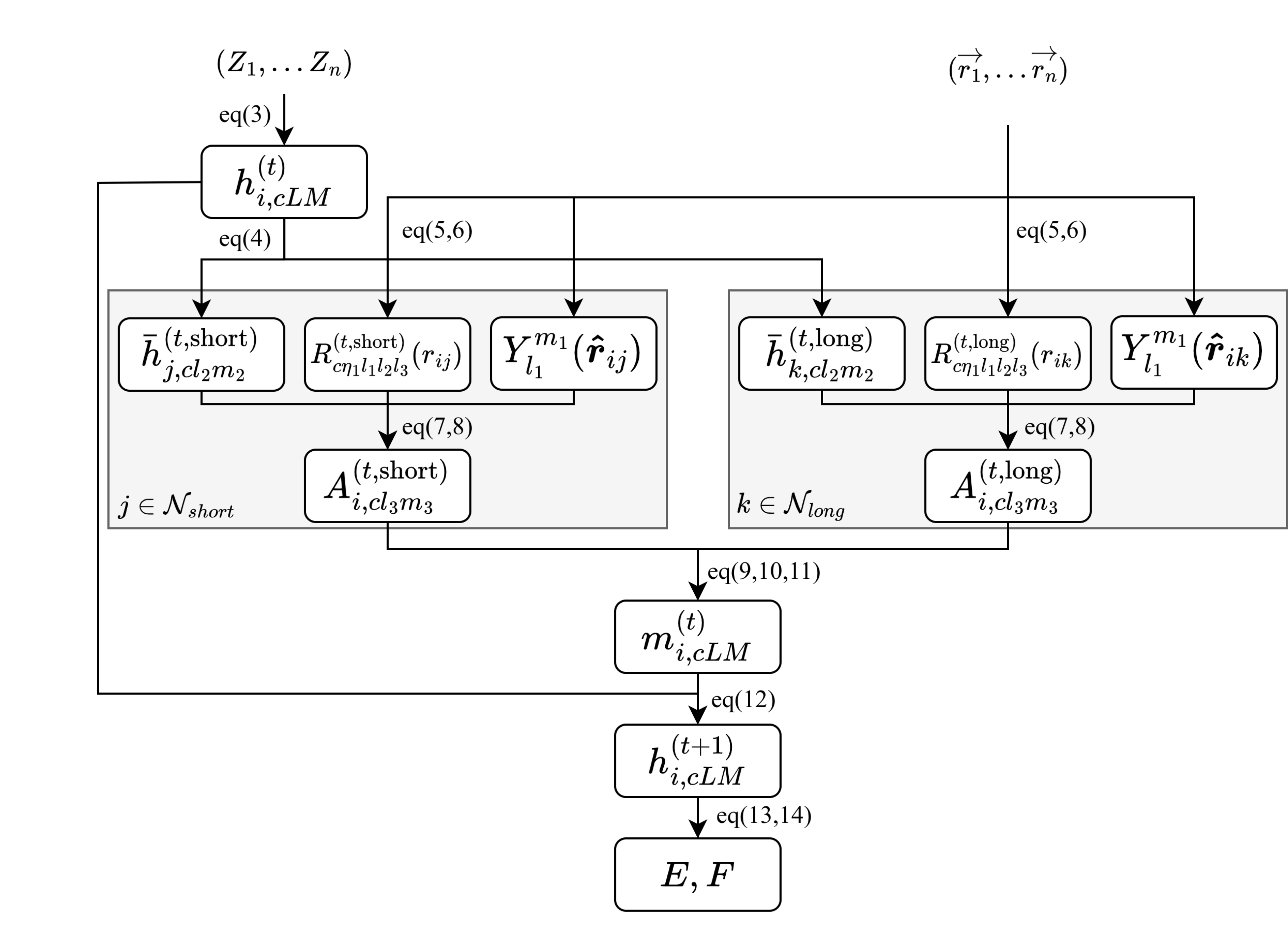}
\caption{\textbar\textbf{Equation calculation flow diagram.}A single interaction layer calculation is completed from \( h^{(t)}_{i,c LM} \) to \( h^{(t+1)}_{i,c LM} \). The final energy is obtained by summing the linearly read-out results from each layer's \( h^{(t+1)}_{i,c LM} \).The forces come from the negative gradient of the final energy with respect to the coordinates.} \label{fig:eq} 
\end{figure*}

Eq. (\ref{eq:phi-basis-t}) is the key part of the equivariant network, merging neighboring atomic features ${h}^{(t,\text{short/long})}_{j,cl_2m_2}$ with the radial information $R_{c \eta_{1} l_{1}l_{2} l_{3}}^{(t,\text{short/long})}(r_{ij})$ and directional information $ Y^{m_{1}}_{l_{1}}$ through convolutional filtering operations. Eq. (\ref{eq:atomic-basis-t}) involves pooling operations and linear transformations. It is worth noting that the handling of long-range information modules may require zero-padding to align with short-range node information, facilitating the final summation in Eq. (\ref{eq:add}).

Eqs. (\ref{eq:symmbasis_L1}) and (\ref{eq:message}) outline the process of efficiently calculating many-body interactions in the MACE framework. Details can be found in the literature. \cite{batatia2022mace,kovacs2023mace-eval} Eq. (\ref{eq:update}) involves residual connections \cite{he2016deep}, designed to update node features. Fig. (\ref{fig:eq})  illustrates the calculation process.

\subsection*{Hyperparameter settings}\label{subsecM3}

All models were trained on a NVIDIA RTX 4090 GPU in single-GPU training using float32 precision.
Unless explicitly stated, the default hyperparameter settings for all models in this paper are as follows: the embedding dimension of the radial basis function is 8; smooth truncation is set to be a polynomial envelope function with p=6; the radial MLP is [64, 64, 64]; the dimension of the readout layer is 16; node features are represented as \texttt{64x0e}; the number of layers in the interaction layer is 2; directional information is expanded to the 3rd order; and the cutoff radius is set to 8 \AA. 

For multiscale higher-order equivariant models, the hyperparameter settings for the short-range node features, the expansion order of short-range directional information, and the number of interaction layers remain the same as described above. However, the long-range node features are represented as \texttt{8x0e} and the long-range directional information is expanded to the first order. The cutoff radius for short-range and long-range are set to 3 \AA\ and 8 \AA, respectively.
For the Allegro model, the env\_embed\_multiplicity is set to 8, and latent\_mlp\_latent\_dimensions is set to 256.

The training hyperparameters include an initial learning rate of 0.01 and a ReduceLROnPlateau scheduler, which reduces the learning rate when the validation loss does not improve over a certain number of epochs. 
To update the evaluation and final model weights of the validation dataset, an exponential moving average with a weight of 0.99 is applied. The optimizer is Adam, and the total number of training epochs is set to 4000. Energy is normalized by the moving average of the potential energy. The loss function is as follows:

\begin{align}
  \mathcal{L} = &
  \frac{\lambda_{E}}{B} \sum_{b=1}^{B} \left(\frac{E_{b} - \hat{E}_{b}}{N_{b}} \right)^{2} \nonumber \\
  & +\frac{\lambda_{F}}{3B} \sum_{b=1}^{B} 
  \frac{1}{N_{b}} 
  \sum_{i_{b},\alpha=1}^{N_{b}, 3} 
  \left(- \frac{\partial E_{b}}{\partial r_{i_{b},\alpha}} - \hat{F}_{i_{b},\alpha} \right)^{2}  \label{eq:loss_fn_old}
\end{align}

The initial weight values of the force and energy follow common settings \cite{batatia2022mace,kovacs2023mace-eval}, where the force weight is set to 1000, and the energy weight is the number of atoms of the system. For MACE and MS-MACE models, Stochastic Weight Averaging (SWA) \cite{izmailov2018averaging, athiwaratkun2018there} is enabled at 75\% of the total iteration count. 
After initiating SWA, as in \cite{kovacs2023mace-eval}, the energy weights and force weights of the loss function are reset to 1000 and 10, respectively.

\subsection*{DFT settings}\label{subsecM4}

Since this paper requires calculations with thousands of atoms and the need to perform AIMD, the energy and forces of all data sets are calculated using the lower computational cost but good accuracy method, i.e., the semi-empirical GFN2-xTB level of theory\cite{bannwarth2019gfn2}. This choice is made to balance computational cost and accuracy. The convergence level follows the default settings of the \texttt{xtb} software \cite{bannwarth2021extended}.

It is essential to note that in the process of data collection, a pretrained MLFF based on the multiscale higher-order equivariant model is used to perform high-precision MD, which is significantly faster than the AIMD simulation. Therefore, the data sampling stage is not computationally intensive. In practice, the primary cost of generating data labeling still lies in labeling using DFT. More accurate quantum-mechanical calculations can be employed based on specific requirements.

\subsection*{Molecular dynamics settings}\label{subsecM5}

Periodic AIMD simulations based on the semi-empirical GFN2-xTB level of theory are implemented using the \texttt{DFTB+} software \cite{hourahine2020dftb}. 
The MD simulations based on the MLFF models are executed using version 1.0 of the \texttt{OpenMM-Torch} plugin and version 8.0.0 of the \texttt{OpenMM} software.

\subsection*{Test of simulation speed and memory consumption}\label{subsecM6}

Before benchmarking, it is essential to preheat the GPU using partial data. The timer employs \texttt{torch.cuda.Event()}, and the dataset with different numbers of atoms should contain at least 20 samples. Each sample is run 50 times, and the average value is taken as the final result. Similarly, \texttt{torch.cuda.max\_memory\_allocated()} from the official torch is used to record the peak GPU memory consumption. Before each recording round, \texttt{torch.cuda.reset\_peak\_memory\_stats()} and \texttt{torch.cuda.empty\_cache()} are employed to reset information and release excess cache. The GPU of the test platform is a 40G A100.

\section*{Code availability}
The code and sample scripts will be released after review.

\section*{Data availability}
The dataset will be released after review.

\section*{Competing interests}
The authors declare no competing interests.

\section*{Acknowledgements}
This work was supported by the National Key R\&D Program of China(No. 2021YFB3803200) and the National Natural Science Foundation of China under Grant No. 22273112.

\bibliography{sn-bibliography}

\end{document}


\section{Local test}

\begin{figure}[h]
\centering
\includegraphics[width=1.0\linewidth]{FigureS1.pdf}
\caption{$\vert$ 
\textbf{a} and \textbf{b} are the changes of the average absolute errors of the energy and forces per atom with the increase of the
atom numbers, respectively. 
\textbf{c} and \textbf{d} are the changes of time cost and memory consumption of simulations with the increase
of atom numbers. The straight lines are obtained by linear fittings to the data points using the least squares method. 
The blue, green, yellow, red, and purple lines are 5 \AA, 6 \AA , 7 \AA , 8 \AA, and 9 \AA, respectively.
}
\label{fig:Local-test}
\end{figure}

Fig. \ref{fig:Local-test} shows the prediction accuracy, simulation speed, and GPU memory consumption of the MACE model  on the perfluorotri-n-butylamine system for different cutoff radius. Energy and force predictions are measured by averaging the mean absolute error (MAE) of models trained with five different random seeds to mitigate the effects of random seed variations.
The results indicate that a cutoff radius of 8 \AA\ achieves optimal accuracy, but results in higher simulation speed and GPU memory consumption.
Overall, with the increase in cutoff radius $r_c$ , the simulation speed and memory
consumption of the model follow a cubic relationship.